\documentclass[aps]{revtex4}

\usepackage[dvipdfm,unicode]{hyperref}
\usepackage[dvipdfm]{hyperref}
\usepackage{graphicx}

\usepackage{verbatim}
\usepackage{amsfonts}
\usepackage{mathrsfs}

\begin{document}

\title{Charmed baryon in a string model  }

\author{Wang Qing-Wu$^{1,2}$ \footnote{qw.wang@impcas.ac.cn} and Zhang Peng-Ming$^{1}$}

 \address{$^{1}$Institute of Modern Physics, Chinese Academy of
Science, P.O. Box 31, Lanzhou 730000, P.R.China}

\address{$^{2}$Graduate School, Chinese Academy of Science,
Beijing 100049, P.R.China}

\begin{abstract}

Charm spectroscopy has studied under a string model. Charmed
baryons are composed of diquark and charm quark which are
connected by a constant tension. In a diquark picture, the quantum
numbers $J^P$ of confirmed baryons are well assigned. We give
energy predictions for the first and  second orbital excitations.
We see some correspondences with the experimental data. Meanwhile,
we have obtained diquark masses in the background of charm quark
which satisfy a splitting relation based on spin-spin interaction.
\end{abstract}

 \maketitle


%%Key words: charm, baryon, diquark, spectroscopy

\section{Introduction  }

 Charm spectroscopy has revived since 2000. Many new excited
charmed baryon states have been discovered by CLEO, BaBar, Belle
and Fermilab. Masses of ground states as well as many of their
excitations are known experimentally with rather good precision.
However charmed baryons have narrow widths and non of their spin
or parity are measured  except the $\Lambda_c(2880)$\cite{Mizuk} .
The assignments listed in the PDG book almost are  based on quark
model\cite{pdg}. Theoretically, the study of heavy baryons has a
long story\cite{Copley,Capstick,Glozman}. Heavy baryons provide a
laboratory to study the dynamics of the light quarks in the
environment of heavy quark, such as their chiral
symmetry\cite{cheng}. The studies of heavy baryons also help us to
understand the nonperturbative QCD\cite{Roberts}. Furthermore, it
really is an ideal place for studying the dynamics of diquark.

The concept of diquark appeared soon after the original papers on
quarks\cite{Gellmann,Ida,Lich}. It was used to calculate the
hadron properties.  In heavy quark effective theory, two light
quarks often refer to as diquark, which is treated as particle in
parallel with quark itself. There are several phenomenal
manifestations of diquark: the $\Sigma-\Lambda$ mass difference,
the isospin $\Delta I=1/2$ rule, the structure function ratio of
neutron to proton, \emph{et al.}
\cite{Anselmino,Jaffe,Selem,Wilczek}. In this diquark picture,
charmed baryons are composed of one diquark and one charm quark.

 Selem and Wilczek\cite{Selem,Wilczek} have generalized the famous
Chew-Frautschi formula by considering diquark and quark connected
by a relativistic string with constant tension $T$ and rotating
with angular momentum $L$. The string is responsible for the color
confinement and is also called as loaded flux tube if the two ends
get masses. In the limit of zero diquark and quark masses, the
usual Chew-Frautschi relationship $E^2 \sim L$ appears. They have
investigated the $N-\Delta$ spectrum and concluded that ``large
$L$ spectroscopy would give convincing evidence for energetically
significant diquark correlations" \cite{Selem}. In hadrons
containing one heavy quark, diquark ought to be a better
approximation than in hadrons containing only light quarks.
However Selem and Wilczek  have given only a short discussion on
the $\Lambda_c$ spectrum. In this paper, we accept the diquark
concept and use their relativistic string model to study the
charmed baryon spectroscopy.

In the following, we introduce firstly  the diquarks and the
string model in section \ref{model}. We give our analysis of the
doublets and give the quantum number assignments based on diquark
assumption. Numerical results are showed in section
\ref{numerical}. In the end summary and discussion are given.

\section{ Diquark and String Model}

\label{model}

\subsection{Diquarks and mass splitting}

 The single-charmed baryons composed of one diquark and one charm quark.
 In literatures, there are two kinds of diquarks: the good
diquark with spin zero and the bad diquark with spin one. The good
diquark is more favorable energetically than the bad one, which is
indicated by both the one-gluon exchange and instanton
calculations. In $SU(3)_f$, the good diquark has flavor-spin
symmetry $\bar {\textbf{3}} _F  \bar {\textbf{3}} _S $ while the
bad diquark $\textbf{6} _F \textbf{6} _S$. To give a color singlet
state, both kinds of diquark have the same color symmetry $\bar
{\textbf{3}}_C $. In the following we use the $[qq^\prime]$ to
denote a good diquark, while $(qq^\prime)$ the bad and $l$ to
denote either  $u$ quark or $d$ quark.

In this diquark picture, $\Lambda_c$ has a $[ud]$ component, while
$\Sigma_c$, $(ll^\prime)$ and $\Omega_c$, $(ss)$. For $\Xi_c$,
either kind of diquarks, good or bad, can be formed. Heavy baryons
always have been obtained by continuum production\cite{Roberts}.
So, the lowest baryons discovered are more likely to be ground
states. These are $\Lambda_c(2285)$ and $\Xi_c(2470)$. Three
doublets $\Sigma_c(2455,2520)$, $\Xi_c(2578,2645)$ and
$\Omega_c(2768,2698)$ would also be states with $L=0$, if they are
composed of bad diquark and charm quark. And we assign the
doublets $\Lambda_c(2595,2628)$ and $\Xi_c(2790,2815)$ with $L=1$.

The good diquark with spin zero has no spin interaction with the
charm quark. So, the lowest energy is a singlet. Only the $L-S$
coupling may make the energy split\cite{Jaffe}:
\begin{equation}
\mathcal{H}(q_c,[qq^\prime])=\mathcal{K}_{[qq^\prime]}2\vec{L}\cdot
\vec{S_c},
\end{equation}
where the coefficient $\mathcal{K}_{L,(qq^\prime)}$ depends on the
diquark and charm quark masses. This interaction splits baryon
with orbital angular momentum L to baryons with $J=L+1/2 $ and
$L-1/2$. And the parity is $P=(-1)^L$.
 For the bad diquark, the
spin-spin interaction is:
\begin{equation}
\mathcal{H}(q_c,(qq^\prime))=\mathcal{G}_{(qq^\prime)}2\vec{S}_{(qq^\prime)}\cdot
\vec{S}_c,
\end{equation}
where $\vec{S}_{(qq^\prime)}$ is the spin of the bad diquark, and
the coefficient $\mathcal{G}_{(qq^\prime)}$ depends on the diquark
and charm quark masses. This spin-spin interaction also lead to a
doublet in the spectrum.  Since in our assignments, there is no
$L>0$ multiplet of bad diquark and for simplicity, we will not
discuss the splitting caused by $L-S$ coupling for baryons
containing bad diquark.

We have relation $<2\vec{j}_1\cdot
\vec{j}_2>=J(J+1)-j_1(j_1+1)-j_2(j_2+1)$, with
$\vec{J}=\vec{j}_1+\vec{j}_2$. It is easy to deduce the mass
difference of a doublet. For example, when $j_1=1$, $j_2=1/2$,
they are $M_0+\Delta$ and $M_0-2\Delta$, with $\Delta$ being
$\mathcal{G}$ or $\mathcal{K}$. Taking a doublet as input, we can
obtain the $\Delta$ and $M_0$. And it is not the masses of the
doublet, but this $M_0$ which enter into the string model.

 \subsection{The string model}

In 1960 Chew and Frautschi conjectured  that the strongly
interacting particles  fall into families where the Regge
trajectory functions were straight lines: $E^2=\sigma+kL$ with the
same constant $k$ for all the trajectories. The straight-line
Regge trajectories with $\sigma$ zero were later understood as
arising from massless endpoints on rotating relativistic strings
at speed of light transversely. A non-zero values of $\sigma$ may
include zero-point energy for string vibrations and loaded
endpoints.

In Selem and Wilczek's model\cite{Selem,Wilczek}, the two ends of
the string have masses $m_1$ and $ m_2$ respectively, with
constant string tension $T$. The rotaing angular momentum is $L$
with angular velocity $\omega$. If the diquark and charm quark are
away from the center of rotation at distances $R_1$ and $R_2$, the
energy of the system is:
\begin{equation}
E=\sum_{i=1,2}(m_i\gamma_i+ \frac{T} {\omega}\int_0^{\omega R_i}
{\frac{1}{\sqrt{1-u^2}}}du), \label{eq:E}
\end{equation}
where $\gamma_i$ is the usual Lorentz factor:
\begin{equation}
\gamma_i=\frac{1}{\sqrt{1-(\omega R_i)^2}}.
 \label{eq:gamma}
\end{equation}
The angular momentum can be written as:
\begin{equation}
L=\sum_{i=1,2}(m_i\omega
R_i^2\gamma_i+\frac{T}{\omega^2}\int_0^{\omega
R_i}{\frac{u^2}{\sqrt{1-u^2}}}du). \label{eq:L}
\end{equation}
Furthermore, we have formula relating the tension
 and the angular velocity:
\begin{equation}
m_i \omega^2 R_i = \frac{T}{\gamma_i ^2}.
 \label{eq:T}
\end{equation}

From equation (\ref{eq:T}), we see that $\omega R_i$ can be
replaced by $m_i$, $\gamma_i$ and $T/\omega$. Solving the
equations (\ref{eq:gamma}) and (\ref{eq:T}), we can express the
$\gamma_i$ by $T/\omega$ and $m_i$:
\begin{equation}
\gamma_i=\sqrt{\frac{1}{2}+\frac{\sqrt{1+4(T/m_i\omega)^2}}{2}}
\end{equation}
Now, we have two equations, (\ref{eq:E}) and(\ref{eq:L}), and 6
parameters, $E$, $L$, $m_1$, $m_2$, $T$ and $\omega$. These
equations are more useful than the Chew-Frautschi formula for they
make us able to extract the diquark masses.

For very light mass, it appears that $\omega \to \infty$ as $L \to
0$ and the Chew-Frautschi relationship  is recovered as $E^2=(2\pi
T)L$. For other cases, such as the first corrections at small
masses  see Ref.\cite{Selem,Wilczek}. A reduced formula has being
used to study the charmed meson spectroscopy\cite{Shan}.In this
paper we will solve the equations numerically.

\section{Numerical Results}
\label{numerical}
\subsection{$\Lambda_c$ and $ \Xi_c$ with good diquark}

\begin{figure}

  \begin{center}
  \includegraphics[width=3in]{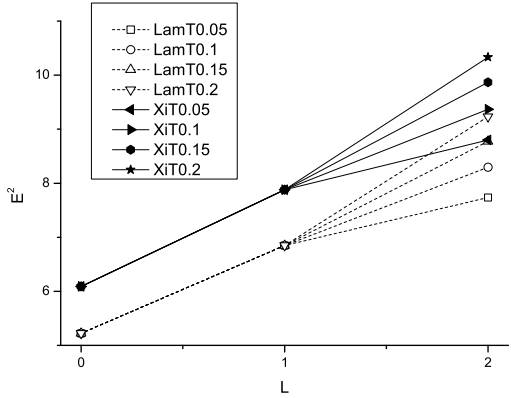}%
  \end{center}
  \caption{Plot of $E^2\sim L$ for $\Lambda_c$ and $\Xi_c$ with $m_c=1.7$ $GeV$ and $T=0.05\sim0.20$. }\label{fig:EL}
\end{figure}

\begin{figure}
  \includegraphics[width=3in]{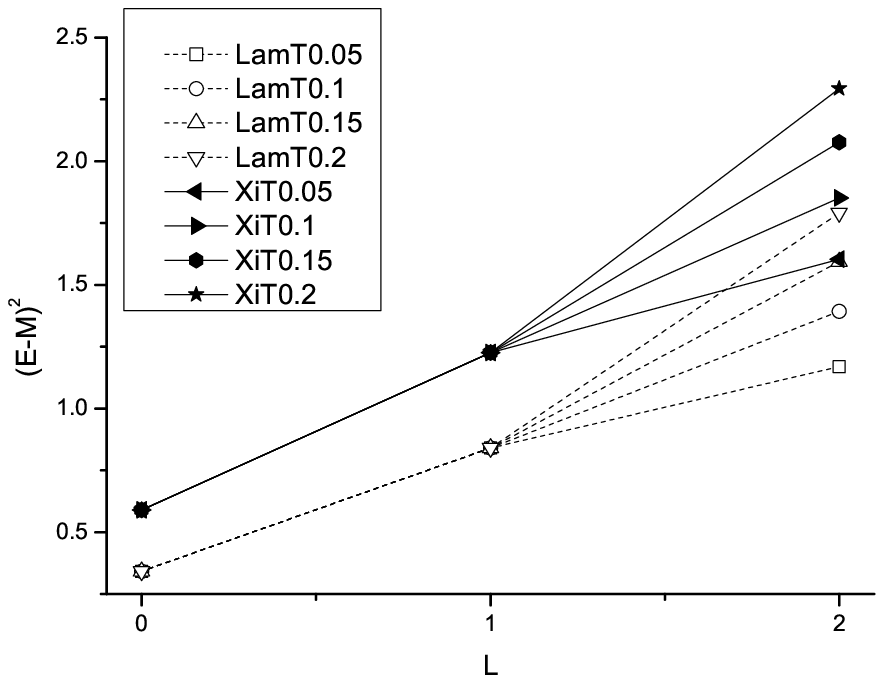}
  \caption{Plot of $(E-M)^2\sim L$ for $\Lambda_c$ and $\Xi_c$ with $m_c=1.7$ $GeV$ and $T=0.05\sim0.20$. }\label{fig:EML}
\end{figure}

We chose $L=1$ doublet as input to solve the equations
(\ref{eq:E}) and (\ref{eq:L}). Firstly, we take the $m_c$ and $T$
as free parameters to get the diquark mass. The string tension $T$
is universal for baryons with the same components. Then we use
these three parameters to give energy predictions for $L=2$.  And
we find that the $T$ is almost equal for $\Lambda_c$ and $\Xi_c$
if we choose the value which give a linear Regge trajectory
$E^2\sim L$ or linear $(E-M)^2\sim L$. The last relation was given
by Selem and Wilczek, which can be obtained by expanding the right
hands of equations (\ref{eq:E}) and (\ref{eq:L}) in $m \omega / T$
for terms of light diquark and in $T/(m \omega)$ for terms of
charm quark. The results are listed in Table \ref{Tab:Lambda} and
\ref{Tab:Xi}. The two kinds trajectories are both linear since the
energies for $L=0,1,2$ at $T=0.1$ form a arithmetic progression
and with small common difference. We plot the two kinds
trajectories with $M_c=1.7$ $GeV$ for example, on Figure
\ref{fig:EL} and \ref{fig:EML}.

%%%%%%%%%%%%%%%%%%%%%%%%%%%%%%%%%%%%%%

%%%%%%%%%%%%%%%%%%%%%%%%%%%%%%%%

\begin{table}
\begin{center}
\begin{tabular}{c|c |c}
\hline\hline $\Lambda_c$: $m_{[ll\prime]}$, $M_0^{L=2}$  & T=0.05
& 0.1
\\
\hline
 Mc=1.5 &0.873,  2.781 &0.737,   2.878
\\
\hline
 1.6&0.770 ,  2.781 &0.630,    2.878
\\
\hline
 1.7&0.665, 2.781& 0.518,   2.879
\\
\hline
 1.8&0.559, 2.782& 0.403,   2.881
 \\
\hline \hline

\end{tabular}
\end{center}

\begin{center}
\begin{tabular}{c|c | c}
\hline\hline
 $\Lambda_c$: $m_{[ll\prime]}$, $M_0^{L=2}$ &   T= 0.15 &   0.2
\\
\hline
 Mc=1.5 &0.624,   2.959& 0.520,    3.032
\\
\hline
 1.6 &0.509,   2.960 &0.395,   3.034
\\
\hline
 1.7& 0.387,   2.962& 0.255,   3.038
\\
\hline
 1.8 &0.253,   2.967&0.080,    3.043
 \\
\hline \hline

\end{tabular}
\end{center}

\caption{Good diquark mass of $[l l^\prime]$ and energy for
$\Lambda_c(L=2)$ neglecting the spin-spin interaction. Mass unit
is in GeV.} \label{Tab:Lambda}
\end{table}

%%%%%%%%%%%%%%%%%%%%%%%%%%%%%%%%%%%%%

%%%%%%%%%%%%%%%%%%%%%%%%%%%%%%%%

\begin{table}
\begin{center}
\begin{tabular}{c|c| c}
\hline\hline
 $\Xi_c$: $m_{[ls]}$, $M_0^{L=2}$& T=0.05 & 0.1
\\
\hline Mc=1.5 & 1.070, 2.968 & 0.940,    3.063
\\
\hline
 1.6 & 0.969,  2.967& 0.835, 3.062
 \\
 \hline
1.7 & 0.866,   2.966 & 0.730,  3.061
\\
\hline
 1.8 & 0.762, 2.967& 0.620, 3.063
\\ \hline\hline
\end{tabular}
\end{center}
\begin{center}
\begin{tabular}{c|c| c}
\hline\hline
 $\Xi_c$: $m_{[ls]}$, $M_0^{L=2}$& T=0.15& 0.2
 \\
 \hline
 Mc=1.5&0.835,   3.142 &0.738, 3.214
 \\
 \hline
1.6& 0.726, 3.141& 0.625, 3.213
\\
\hline 1.7 & 0.614, 3.141& 0.505, 3.214
\\
\hline 1.8 & 0.498,  3.143& 0.380, 3.216
\\
\hline\hline
\end{tabular}
\end{center}
\caption{Good diquark mass of $[l s]$ and energy for $\Xi_c(L=2)$
neglecting the spin-spin interaction. Mass unit is in GeV.}
\label{Tab:Xi}
\end{table}

The energies $M_0^{L=1}$ is $2.617$ $GeV$  for doublet
$\Lambda_c(2595,2628)$ with $\mathcal{K}_{[ll]}=11$ $MeV$. The
numerical result for $L=2$ are $M_0=2.879$ $GeV$ which gives
doublets $\Lambda_c^\prime(2846,2901)$  with the splitting mass
formulas $M_0+2\mathcal{K}$ and $M_0-3\mathcal{K}$. Here, we use a
prime to indicate our theoretic prediction. We think the
$\Lambda_c(2880)$ and  $\Lambda_c(2940)$ to be a doublet with
$L=2$, since the splitting gives $\mathcal{K}_{[ll]}=12$ $MeV$
which is near equal to the result from  mass difference of
$\Lambda_c(L=1)$. Then their mass difference gives
$M_0^{L=2}=2917$ $MeV$ which is 38 $Mev$ large than our prediction
However, it can be wiped out given $T=1.2$, see Table
\ref{Tab:gT0.12}.

The linear fit is:
 \begin{equation}
\Lambda_c:  E^2=1.632L+5.220,
 \end{equation}
 with $\chi/Dof$ almost being zero.
 So we predict that the quantum numbers $J^P$ of doublet
$\Lambda_c(2882,2940)$ are $3/2^+$ and $5/2^+$.

For $\Xi_c$, $M_0^{L=1}$ is $ 2.807$ $GeV$ with
$\mathcal{K}_{[ls]}=8.3$ $MeV$. If $T=0.12$, we get
$M_0^{L=2}=3.094$ $GeV$ and the doublet is
$\Xi_c^\prime(3069,3111)$. And the Regge trajectory is
  \begin{equation}
\Xi_c:  E^2=1.736L+6.115.
 \end{equation}
 The nearest
experimental data are $\Xi_c(3080)$ and $\Xi_c(3123)$ only about
$10 MeV$ larger than our predictions. So, we take
$\Xi_c(3080,3123)$ as a doublet with $J^P=$ $3/2^+$ and $5/2^+$.

%%%%%%%%%%%%%%%%%%%%%%%%%%%%%%%%%%%%%TTTTT
\begin{table}
\begin{center}
\begin{tabular}{c|c c  }
\hline \hline $B_{(qq\prime)}$& $\Lambda_c$ & $\Xi_c$
\\
\hline
 $\mathcal{K}$/MeV& 11& 8.3
\\
\hline
 $m_{[qq\prime]}$ /GeV&0.465&0.682
\\
\hline
 $M_0^{L=1}/GeV$ &2.617 & 2.807
\\
\hline
 $M_0^{L=2}/GeV$ & 2.913 &3.094
\\
\hline mass splitting & L=1: $M_0+ \mathcal{K}$ and $M_0-2\mathcal{K}$\\
\cline{2-3} & L=2:  $M_0+2\mathcal{K}$ and $M_0-3\mathcal{K}$

\\ \hline\hline

\end{tabular}
\end{center}
\caption{Good diquark masses and predictions for masses at $L=2$
with $T=0.12$ and $m_c=1.7$ $GeV$. By using the mass splitting
formula at $L=1$, $M_c^{L=1}$ and $\mathcal{K}$ are easy to be
derived. } \label{Tab:gT0.12}
\end{table}
\subsection{$\Sigma_c$, $\Xi_c$ and $\Omega_c$ with bad diquark}

For $\Sigma_c$, $\Omega_c$ and $\Xi_c$ with bad diquark, we have
to take the $L = 0$ doublets as input for the lack of data. When
$L \to 0$, we have $\omega \to 0$, $R \to 0$ and $E \to m_1+m_2$
from which we can deduce the bad diquark masses. We see from Table
\ref{Tab:Lambda} and Table \ref{Tab:Xi} that energies are more
depending on $T$ not on quark masses. So  we take charm quark mass
to be $1.7$ $GeV$. The numerical results with $T=0.12$ are listed
in Table \ref{Tab:baddiq}. Linear fits of the three groups of
baryon masses are:
\begin{eqnarray*}
\Sigma_c:   E^2=1.987L+6.326,\\
\Xi_c:   E^2=2.035L+6.967,\\
\Omega_c:   E^2=2.084L+7.624,
\end{eqnarray*}
with $\chi/Dof$ being about $0.04$ for each fit. The slopes are
almost equal but a little larger than 1.632 and 1.736, the slopes
for fitting the spectra of $\Lambda_c$ and $\Xi_c$ containing good
diquark. However, the  diquark masses are so heavy. And it is
unreasonable for a string with zero length. So, when $L\to0$ and
$R\to0$, the string model would not be a good approximation.

In the end, we give the mass predictions for these baryons using
linear Regge trajectory, though there are arguments that hadronic
Regge trajectories are  nonlinear\cite{nonlinear}. We take the
slope to be the average of the slopes for good diquark baryons,
that is 1.684. Then use equations (\ref{eq:E}) and (\ref{eq:L})
with L=2 to extract the diquark masses. Results are showed in
Table \ref{Tab:slope}.  In PDG book, there is $\Sigma_c(2800)$
with question mark which is a little lower than our prediction for
$\Sigma_c(2815,L=1)$. And note that we have neglected here all the
angular momentum interactions.

%%%%%%%%%%%%%%%%%%%%%%%%%%%%%%%%%%%%%TTTTT
\begin{table}
\begin{center}
\begin{tabular}{c|c c  c}
\hline \hline $B_{(qq\prime)}$& $\Sigma_c$ & $\Xi_c$ & $\Omega_c $
\\
\hline
 $\mathcal{G}$/MeV&  21.7& 22.3&23.3
\\
\hline
 $m_{(qq\prime)}$ /GeV&0.798&0.923&1.045
 \\
\hline
 $M_0/GeV$& 2.498 & 2.623 &2.745
\\
\hline
 $M_0^{L=1}/GeV$ &2.913 & 3.029& 3.144
\\
\hline
 $M_0^{L=2}/GeV$ & 3.196 &3.309 &3.421

\\ \hline\hline

%%%%%%%%%%%%%%%%%%%%%%%%%%%%%%%%%%%%%%%%%%%%%%%%%%
\end{tabular}
\end{center}
\caption{Results of the string model using the  ground states as
input and with $T=0.12$.} \label{Tab:baddiq}
\end{table}
%%%%%%%%%%%%%%%%%%%%%%%%%%%%%%%%%%%%%TTTTT
\begin{table}
\begin{center}
\begin{tabular}{c|c c  c}
\hline \hline $B_{(qq\prime)}$& $\Sigma_c$ & $\Xi_c$ & $\Omega_c $
\\
\hline
 $\mathcal{G}$/MeV&  21.7& 22.3&23.3
\\
\hline
 $m_{(qq\prime)}$ /GeV&0.739&0.858&0.975
 \\
\hline
 $M_0/GeV$& 2.498 & 2.623 &2.745
\\
\hline
 $M_0^{L=1}/GeV$ &2.815 & 2.926& 3.036
\\
\hline
 $M_0^{L=2}/GeV$ & 3.100 &3.201 &3.302

\\ \hline\hline

\end{tabular}
\end{center}
\caption{Results of  using Regge trajectory with slope being
1.684. And the diquark masses are derived by taking the
$M_0^{L=1}$ as input. \label{Tab:slope}}
\end{table}
\subsection{Diquark masses}

The good and bad diquark masses are listed in Table
\ref{Tab:gT0.12} and Table \ref{Tab:slope}. Bad diquarks are
heavier than good diqarks and diquark with heavier flavor quark is
heavier than the light one. These diquark masses are sensitive to
the background, i.e. the charm quark mass. However, they still
satisfy the relation $(ud)-[ud]$ $>$ $(us)-[us]$ which was
expected from spin-spin interaction that the mass difference would
be strongest for lightest quarks\cite{Wilczek,Selem}.

We can adopt the string model to the charmed mesons. The
non-strange mesons $D(2400,2420,2430,2460)$ with positive parity
would be a multiplet of $L=1$. The meson $D(2460,J^P=2^+)$ thus
has  total spin $S=1$ and $<2L \cdot S>=0$. The same is for
charmed and strange meson $D_s(2573)$. Using this two states as
input, we have derived the quark masses which are $0.332$ $GeV$
for up and down quark and $0.468$ $GeV$ for strange quark. The
diquark masses can be defined by $M_D=M_{q1}+M_{q2}+E_{12}$, with
$E_{12}$ being the  binding energy. We see that the good diquarks
have negative  binding energies while the bad positive. This is
consistent with result that comes from spin-dependent
colormagnetic interactions of two quarks. The interactions are
attractive in a spin-0 state and repulsive in a spin-1
state\cite{Lichtenberg}.

 \section{Summary and Discussion}
\label{summary}

We have used a diquark picture and a string model to study the
charmed baryon spectroscopy. The many doublets in the spectroscopy
are the results of S-S or L-S interactions. With string tension
$T=0.12$ we have given predictions for the good diquark baryons
with L=2 which have some experimental correspondences. The
possible state $\Sigma_c(2800)$ would be the first orbital
excitation of $\Sigma_c$. The quantum number $J^P$ assignments for
L=0 and L=1 baryons from a diquark picture are the same as PDG
book. By using the string model, we have extracted the diquark
masses which satisfy the expected relation $(ud)-[ud]$ $>$
$(us)-[us]$.

However, there is one problem. Our prediction for
$\Lambda_c(2880)$ $J^P=3/2^+$ is contradicted with the
experimental result and Selem's assignment with
$J^P=5/2^+$\cite{Mizuk,Selem}. If it is confirmed by later
experiments, we must reconsider our diquark picture or mass
splitting formula based on angular momentum interactions.

\section{Acknowledgments}
This work was supported by  Chinese Academy of Sciences Knowledge
Innovation Project (KJCX2-SW-No16;KJCX2-SW-No2), National Natural
Science Foundation of China(10435080;10575123), West Light
Foundation of The Chinese Academy of Sciences and Scientific
Research Foundation for Returned Scholars, Ministry of Education
of China.

\end{document}